\documentclass[twocolumn,showpacs,preprintnumbers,amsmath,amssymb]{revtex4}

\usepackage{graphicx}
\usepackage{dcolumn}
\usepackage{bm}

\begin{document}


\title{Energy Dependence of Directed Flow in Au+Au Collisions \\from a Multi-phase Transport Model}

\affiliation{Institution of Particle Physics, Huazhong Normal
University(CCNU), Wuhan 430079, P.R.China}

\affiliation{The Key Laboratory of Quark and Lepton Physics
(Huazhong Normal University), Ministry of Education, Wuhan 430079,
P.R.China }

\affiliation{Institute of High Energy Physics, CAS, Beijing
100049,P.R.China }

\affiliation{Shanghai Institute of Applied Physics, CAS, Shanghai
201800, P.R.China}

\affiliation{Physics Department, Brookhaven National Lab., Upton, NY
11973, USA}

\author{J.~Y.~Chen}\email{chenjy@iopp.ccnu.edu.cn}\affiliation{Institution of Particle
Physics, Huazhong Normal University(CCNU), Wuhan 430079, P.R.China}
\affiliation{The Key Laboratory of Quark and Lepton Physics
(Huazhong Normal University), Ministry of Education, Wuhan 430079,
P.R.China}

\author{J.~X.~Zuo}\email{zuojx@ihep.ac.cn}\affiliation{Institute of High Energy
Physics, CAS, Beijing 100049, P.R.China }

\author{X.~Z.~Cai}\affiliation{ Shanghai Institute of Applied Physics, CAS, Shanghai 201800, P.R.China}

\author{F.~Liu}\affiliation{ Institution of Particle Physics, Huazhong Normal
University(CCNU), Wuhan 430079, P.R.China} \affiliation{The
Key Laboratory of Quark and Lepton Physics (Huazhong Normal
University), Ministry of Education, Wuhan 430079, P.R.China}

\author{Y.~G.~Ma}\affiliation{Shanghai Institute of Applied Physics, CAS, Shanghai 201800, P.R.China}

\author{A.~H.~Tang}\affiliation{Physics Department, Brookhaven National Lab., Upton, NY
11973, USA}

\date{\today}

\begin{abstract}
The directed flow of charged hadron and identified particles has been
studied in the framework of a multi-phase transport (AMPT) model,
for $^{197}$Au+$^{197}$Au collisions at $\sqrt{s_{NN}}=$200, 130,
62.4, 39, 17.2 and 9.2 GeV. The rapidity, centrality and energy
dependence of directed flow for charged particles over a wide
rapidity range are presented. AMPT model gives the correct $v_1(y)$
slope, as well as its trend as a function of energy, while it
underestimates the magnitude. Within the AMPT model, the proton $v_1$ slope
is found to change its sign when the energy increases to 130 GeV - a
feature that is consistent with ``anti-flow''. Hadronic
re-scattering is found having little effect on $v_1$ at top RHIC
energies. These studies can help us to understand the collective
dynamics at early times in relativistic heavy-ion collisions, and
they can also be served as references for the RHIC Beam Energy Scan
program.
\end{abstract}

\maketitle

\section{Introduction}

Anisotropic flow is one of the key observables in characterizing
properties of the dense and hot medium created in the relativistic
heavy-ion collisions\cite{ARTpaper}. It is quantified by Fourier
coefficients when expanding particle's azimuthal distribution with
respect to the reaction plane\cite{Methods}:
\begin{equation}
E\frac{d^{3}N}{d^{3}p}=\frac{1}{2\pi}\frac{d^{2}N}{p_{T}dp_{T}dy}(1+\sum_{n=1}^{\infty}2v_{n}\cos
n\phi)
\end{equation}
where $\phi$ denotes the angle between the particle's azimuthal
angle in momentum space and the reaction plane angle. The sine terms
in Fourier expansions vanish due to the reflection symmetry with
respect to the reaction plane. The various coefficients in this
expansion can be defined as:

\begin{equation}
v_{n}=\langle\cos n\phi\rangle
\end{equation}

The first and the second coefficients are named as directed flow
$(v_{1})$ and elliptic flow$(v_{2})$, respectively, and they play
important roles in describing the collective expansion in azimuthal
space. Elliptic flow is produced by the conversion of the initial
coordinate-space anisotropy into momentum-space anisotropy, due
to the developed large in-plane pressure gradient. Elliptic flow
depends strongly on the re-scattering of the system constituents,
thus it is sensitive to the degree of
thermalization\cite{Heinzpaper} of the system at early time.
Directed flow, which is the focus of this study, describes the
``side splash" of particles away from mid-rapidity\cite{Sorge}, and
it probes the dynamics of the system in the longitudinal direction.
Since directed flow is generated very early, it brings
information from the foremost early collective motion of the system.
The shapes of directed flow, in particular those for identified
particles, are of special interest because they are sensitive to the
equation of state (EOS) and may carry a phase transition
signal\cite{Stocker}.

The study of energy dependence of directed flow has implications in
many aspects. Firstly, because directed flow has a unique,
pre-equilibrium origin, it is expected to behave differently than
other soft observables which show an ``entropy-driven" multiplicity
scaling\cite{scalingSoft}. It has been shown by
STAR\cite{STARv1ZDCSMD} that at top RHIC energies, directed flow is
independent of system size, while it has an energy dependence. For a
comprehensive study of the subject, it is necessary to extend the
study of the energy dependence of directed flow in a wider energy range.
Secondly, experiments at RHIC (PHENIX and STAR) have planned
to look for the existence of the QCD phase boundary and the possible
critical point by colliding heavy ions at various incident beam
energies\cite{Paulpaper,STARBES,PHENIXplan}. A non-monotonic
dependence of variables on $\sqrt{s_{NN}}$ and an increase in
event-by-event fluctuations should become apparent near the critical
point\cite{Paulpaper}. Directed flow is generated during the nuclear
passage time $(2R/\gamma \sim 0.1 fm/c)$ and it probes the onset of
bulk collective dynamics in the earlier stage of the collision. As a
suggested signature of a first order phase transition\cite{Stocker},
directed flow is sensitive to the creation of the critical point and
it plays an important role in the proposed beam energy scan program.

In this paper, directed flow from the AMPT model for 6 energies are
presented. They are, 9.2, 17.3, 39, 62.4, 130 and 200 GeV. The
comparison with the measurements from STAR and PHOBOS are made at
top energies. The particle type dependence over a wide rapidity
range is discussed. This study will deepen our understanding about
the energy dependence of directed flow, and it can be also served as
a valuable reference for the RHIC Beam Energy Scan program.

\section{The AMPT model}

The AMPT model consists of four main components\cite{AMPTLongPaper}:
initial conditions, partonic interactions, conversion from partonic
matter to hadronic matter, and hadronic interactions. The initial
conditions, which include the spatial and momentum distributions of
the mini-jet partons and soft string excitations, are obtained from
the HIJING model\cite{HIJING}. The scatterings among partons are
modeled by Zhang's parton cascade(ZPC)\cite{ZPC}, which includes
two-body scatterings with cross sections from pQCD with screening
masses. In the default AMPT model\cite{default}, partons are
recombined with their parent strings when they stop interacting, and
the resulting strings fragment into hadrons according to the Lund
string fragmentation model\cite{Lund}. In the AMPT model with string
melting\cite{stringMelting}, quark coalescence is used instead to
combine partons into hadrons. The dynamics of the subsequent
hadronic matter is described by the ART (a relativistic transport)
model\cite{ART} with modifications and extensions. As suggested in
Ref.\cite{AMPTv1old}, the parton cross section is chosen as 3 mb in
our analysis. All the errors presented here are statistical only.

\section{Analysis and Results}
\begin{figure}
\resizebox{0.48\textwidth}{!}{%
  \includegraphics{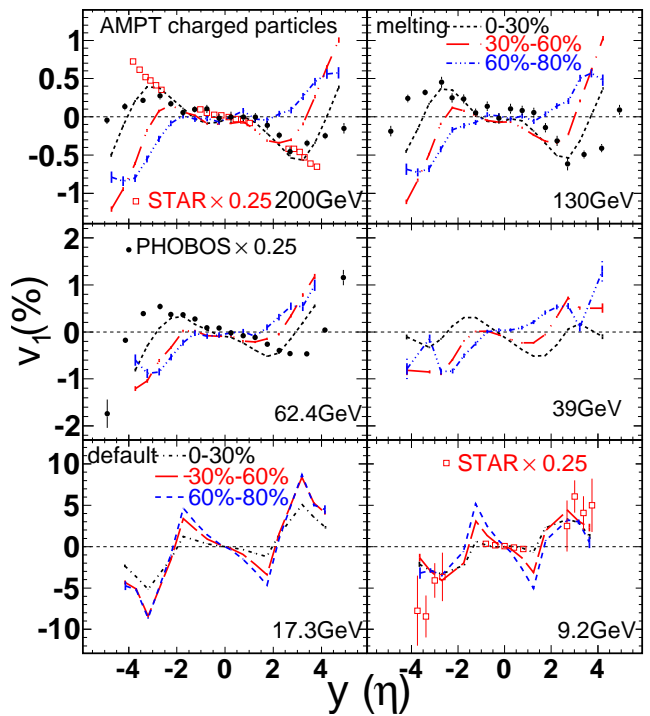}
}
\vspace{0.01cm}       
\caption{Rapidity dependence of $v_{1}$ for charged particles in
the AMPT model compared with STAR and PHOBOS data (plotted as a
function of $\eta$) in the Au+Au collisions at
$\sqrt{s_{NN}}=200$GeV. The dashed lines showed AMPT result from
different centrality: 0-30\%(black), 30\%-60\%(red),
60\%-80\%(blue).}
\label{fig:v1Chg}       
\end{figure}

In Fig.~\ref{fig:v1Chg}, the directed flow of charged particles from
AMPT is shown as a function of rapidity, for collision energies of
200, 130, 62.4, 39, 17.2 and 9.2GeV. The centrality is divided into
three bins, namely, 0-30\%, 30\%-60\% and 60\%-80\%, based on the
impact parameter ($b$) distribution. The calculations with string
melting scenario is used for high energies (200, 130, 62.4, 39GeV)
while for low energies (17.2 and 9.2GeV) calculations are performed
with default scenario.  The reason for such choice is because that,
it is argued~\cite{JinhuiPaper,JiaxuPaper,AMPTv1old} that the string
melting should be used to explain flow around midrapidity at top
RHIC energies, and default setting describes data at 9.2 GeV the
best. The energy density in the collisions at the RHIC top energies
is mush higher than the critical density for the QCD phase
transition. More discussion on different AMPT configurations can be
found later in this paper. All results are obtained by integrating
over transverse momentum ($p_{T}$) up to 4.0 GeV/c. Experimental
results from STAR\cite{STARv1ZDCSMD,LokeshPaper} and
PHOBOS\cite{PHOBOSv1} are also shown for comparison. The charged
hadron $v_{1}$ measured by the PHOBOS experiment is for 0-40\%
central collisions, and the results measured by STAR experiment are
for centrality 30\%-60\% at 200GeV, and centrality 0-60\% at 9.2GeV.
In general, AMPT gives larger $v_1$ at low energies than at high
energies, the same trend has been seen in data. At top RHIC
energies, AMPT underestimated $v_{1}$, due to the turn-off of
mean-field potentials in ART when implemented in AMPT to describe
the hadronic scattering\cite{AMPTLongPaper}. However, in the
rapidity range of [-2.0,2.0], the shape of $v_1$ between AMPT
calculations and experimental data are in good agreement $-$ this
can be seen by scaling experimental results with a factor of 0.25.

\begin{figure}
\resizebox{0.48\textwidth}{!}{%
  \includegraphics{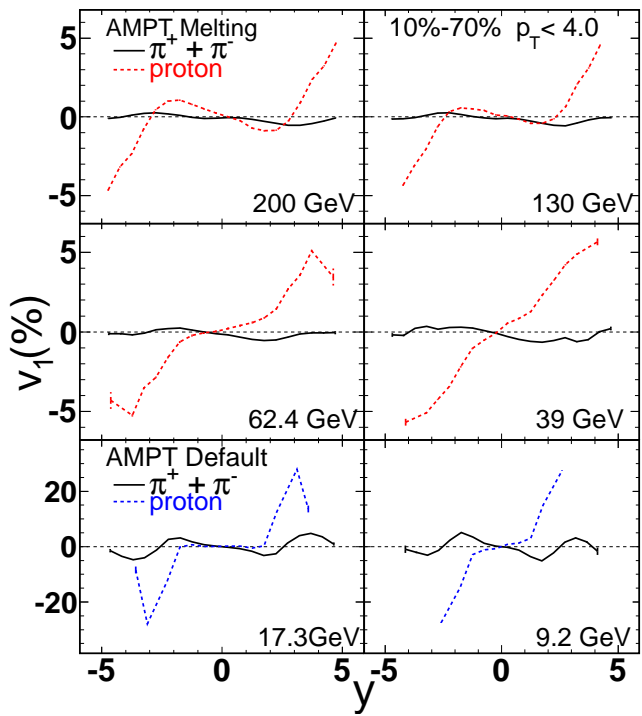}
}
\caption{Proton(solid lines) and Pion(dashed lines) $v_{1}(y)$ from
AMPT at centrality 10\%-70\%.}
\label{fig:v1PID}       
\end{figure}

The particle type dependence of directed flow is shown in
Fig.~\ref{fig:v1PID}. The different sign of $v_1$ between pions and
protons at low energies can be understood as nucleon shadowing and
baryon stopping\cite{Nupaper,Liu1998}. In general the magnitude of
the $v_1$ slope at midrapidity decreases with increasing energy.
This effect is most profound for protons, for which the slope keeps
decreasing and when the energy is high enough, it changes its sign
and protons begin to flow together with pions. This is consistent
with the ``anti-flow'' scenario\cite{antiFlow}, in which the
``bounce-off'' motion and transverse expansion of nucleons compete
with each other around midrapidity, and when the transverse
expansion is strong enough (e.g., at top RHIC energies), it
overcomes the ``bounce-off'' motion and causes protons to change their
sign of directed flow.

\begin{figure}
\resizebox{0.48\textwidth}{!}{%
  \includegraphics{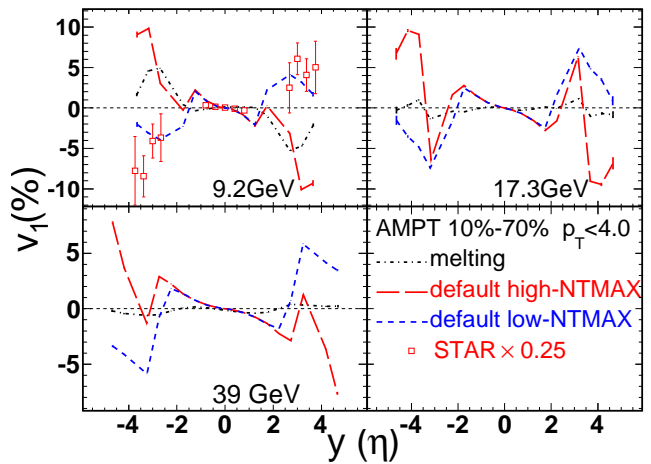}
}
\vspace{0.01cm}       
\caption{Charged particles $v_{1}(y)$ from centrality 10\%-70\% for
9.2GeV(upper left panel),17.3GeV(upper right panel) and 39GeV(down
left panel). The dashed lines show three AMPT versions : string
melting scenario(black), default scenario with high-NTMAX(red) and
low-NTMAX(blue). Experimental data points from STAR are plotted as a
function of $\eta$.}
\label{fig:v1ThreeCases}       
\end{figure}

To illustrate the effect on $v_1$ due to different configurations in
AMPT, in Fig.~\ref{fig:v1ThreeCases} we present the directed flow of
charged hadrons in low energy collisions, from AMPT calculations
with the string melting scenario and the default scenario. The
similar study for higher energies can be found in \cite{AMPTv1old}.
The calculation with string melting yields the smallest $v_1$ slope
around mid-rapidity and is close to data. Two different default
scenarios are also studied: one is calculated with NTMAX=2500
(high-NTMAX), and the other, NTMAX=150 (low-NTMAX). NTMAX stands for
the number of time-steps for the hadron cascade (see detail in paper
\cite{AMPTLongPaper}). A large NTMAX means a thoroughly developed
hadron cascade, as  0.2fm/c*NTMAX is the termination time, in the
center of mass frame, of the hadron cascade in AMPT model. The
comparison, for low energies, of $v_1$ calculated between low-NTMAX
and high-NTMAX indicates that $v_1$ can change its sign at large
rapidity if the time for the hardonic cascade is long enough. In default
AMPT, the NTMAX has to be much larger than 150 in order to describe
$v_1$ at large rapidity. The disagreement between the experimental
data and the calculation made with high-NTMAX is mostly due to the
lack of the mean-field in the hadron cascade in AMPT, which is a
considerable effect at low energies when the nuclei passage time is
not negligible (compared to that at high energies). The AMPT
calculation with high-NTMAX at high energy has been presented in
\cite{AMPT_PLB}. In this paper, we address the comparison around
midrapidity only, and results presented in this paper are made with
low-NTMAX unless otherwise specified.

\begin{figure}
\resizebox{0.48\textwidth}{!}{%
  \includegraphics{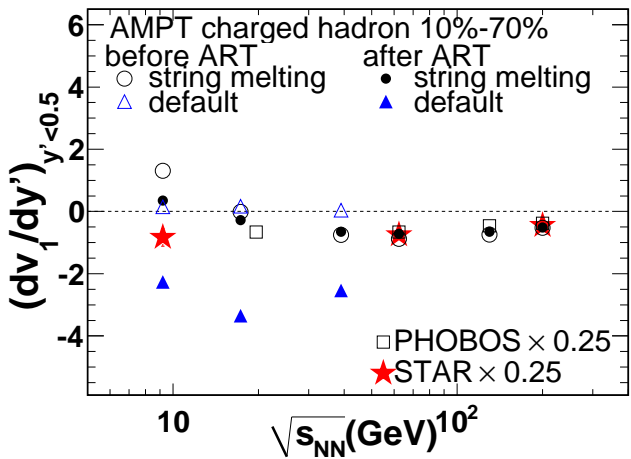}
}
\vspace{0.5cm}
\caption{Charged hadrons' slope $dv_{1}/dy'$ in the
mid-rapidity $|y'|<0.5$ as a function of incident-energy. The data
are taken from STAR(stars) and PHOBOS(squares) and scaled by a
factor 0.25. The AMPT calculatons with string melting before ART are
depicted with open circles and after hadron cascade are depicted
with full circles. The open triangles depict the default AMPT
calculaitons before ART and the full triangles depict after hadron
cascade.}
\label{fig:v1Integrated}       
\end{figure}

The energy dependence of charged particle directed flow, calculated
with the AMPT model, is shown in Fig.~\ref{fig:v1Integrated}.
Experimental data are also shown for comparison. The centrality for
which the calculation is performed is 10\%-70\%. The centrality for
PHOBOS data from different energies is 0-40\% while the centrality
selection for STAR data are 0-60\% for 9.2 GeV, 10\%-70\% for 62.4
GeV, and 30\%-60\% for 200 GeV. To obtain the integrated $v_1$, one
needs to fold in the spectra at different energies, which brings in
an additional layer of systematics. Thus instead, we present the slope
of $v_{1}(y)$ around mid-rapidity ($|y'|<0.5$) extracted from the
normalized $(y'=y/y_{beam})$ rapidity distribution, where $y_{beam}$
is the beam rapidity. For the energy range that string melting is
used (39 GeV and above), all the AMPT calculations underestimate the
experimental data, however, they predict the right trend of the
energy dependence. For the low energies at 9.2 GeV, calculations
with string melting did a poor job, the calculation with the default
AMPT improves the result in the right direction yet is still not be
able to explain the data. The hadron re-scattering effect on
directed flow $v_{1}$ can be seen by switching off the hadron cascade in
the AMPT calculation. Comparing the difference between the result with
hadron cascade (open symbols) and without (solid symbols), it is
found that the hadronic cascade has a significant effect for low energy
results but little for that of high energies. This can be understood
as that, when the energy is high enough, the hadron re-scattering
become less important due to the presence of strong collective
motion built up beforehand.

\vspace{6pt}
\section{Summary and Conclusions}

In this paper, $v_1$ values calculated from the AMPT model for different
energies are discussed. It is found that the AMPT model gives the right
shape of $v_1$ versus $y$ while underestimating the magnitude,
possibly due to the lack of mean-field in its hadron cascade. In
AMPT, the proton $v_1$ slope changes its sign when the energy
increases to 130 GeV and begins to have the same sign as that of
pions, as expected in the ``anti-flow'' scenario. The effect on
$v_1$ due to string melting, low-NTMAX and high-NTMAX are
illustrated. The energy dependence of the $v_1$ slope at midrapidity is
compared to experimental data, and AMPT can describe the trend of
energy dependence while missing the magnitude by a fraction of 75\%.
Hadronic rescattering is found to be less important at
high energies as the strong collective motion becomes to be the
dominant dynamics.

\begin{acknowledgments}
Authors greatly thank Zi-Wei Lin and Zhangbu Xu for useful
discussions and kindly providing comments to this paper. Authors
appreciate Matthew Lamont's help on English QA. This work was
supported in part the National Natural Science Foundation of China
under Grant No. 10775058 \& No. 10610285, MOE of China under Grant
No. IRT0624 and MOST of China under Grant No. 2008CB817707, and the
Knowledge Innovation Project of the Chinese Academy of Sciences
under Grant Nos. KJCX2-YW-A14 and and KJCX3-SYW-N2.
\end{acknowledgments}


\end{document}